\documentclass[%
twocolumn,
superscriptaddress,
nofootinbib,
 amsmath,amssymb,
 aps, physrev,
]{revtex4-2}

\usepackage{graphicx}% Include figure files
\usepackage{dcolumn}% Align table columns on decimal point
\usepackage{bm,xcolor,ulem}% bold math
\usepackage{makecell}
\usepackage{booktabs} % For \toprule, \midrule, \bottomrule

\begin{document}

\preprint{APS/123-QED}

\title{Grain Boundary Diffusion in Yukawa Crystals}

\author{Matthew E. Caplan}
\email{Corresponding author: mecapl1@ilstu.edu}
\affiliation{Department of Physics, Illinois State University, Normal, IL 61790, USA}
\affiliation{Department of Physics, University of Illinois Urbana–Champaign, Urbana, IL 61801, US}

\author{Nevin T. Smith}
\affiliation{Department of Physics, Illinois State University, Normal, IL 61790, USA}

\author{Dany Yaacoub}
\affiliation{Department of Physics, Illinois State University, Normal, IL 61790, USA}
\affiliation{Canadian Institute for Theoretical Astrophysics, 60 St. George Street, Toronto, ON M5S 3H8, Canada}%

\author{Roberto F. Serrano}
\affiliation{Department of Physics, Illinois State University, Normal, IL 61790, USA}
\author{Elias Taira}
\affiliation{Department of Physics, Illinois State University, Normal, IL 61790, USA}

\author{Ashley Bransgrove}
\affiliation{Princeton Center for Theoretical Science and Department of Astrophysical Sciences, Princeton University, Princeton, NJ 08544, USA}

%\collaboration{CLEO Collaboration}%\noaffiliation

\date{\today}% It is always \today, today,
             %  but any date may be explicitly specified

\begin{abstract}
We present calculations of diffusion coefficients in grain boundaries in Yukawa crystals for astrophysics. Our methods follow from our recent work calculating diffusion coefficients in perfect body-centered cubic crystals. These diffusion coefficients show only a weak dependence on the crystal orientations at the grain boundary and are consistent with those expected for a supercooled liquid scaled down by one to two orders of magnitude. We argue that the local disorder at the grain boundary produces a landscape of potential barriers similar to that of an amorphous liquid thin film, significantly reducing activation barriers to diffusive hops relative to the bulk solid. This also introduces a screening dependence, such that boundary diffusion does not exhibit the same universality as the bulk crystal. These diffusion coefficients suggest that grain boundaries may be a dominant source of viscous dissipation in neutron star crusts.  
\end{abstract}

%\keywords{Suggested keywords}%Use showkeys class option if keyword
                              %display desired
\maketitle

%\tableofcontents

\section{Introduction}\label{sec:intro}

Diffusion at grain boundaries (GBs) is an important transport mechanism in polycrystalline materials. In cases where bulk diffusion within a crystallite is slow, for example at low temperatures, GB diffusion can be the dominant form of dissipation. We argue that this is likely to be the case in neutron star crusts. GB diffusion coefficients are essential microphysics input for modeling material behavior during any elastic evolution of the crust, including creep, grain growth, and breaking events involving plastic flow and viscous dissipation. Indeed, because of the high pressure brittle fractures and voids cannot propagate in neutron star crusts \citep{jones2003nature}, so that when the crust breaks, for example in a starquake \citep{Ruderman_1969R, Blaes_1989B, Thompson_1995, Bransgrove_2020}, one might expect the viscous dissipation to be set by the GB diffusion coefficients.

In our recent work, \citet{Caplan2024} (hereafter CY24), we show that diffusion in Yukawa crystals is dominated by the thermally activated formation of vacancy–interstitial pairs. In the absence of these defects, nuclei are trapped on a lattice site by the Coulomb barriers of their nearest neighbors and to diffuse they require a nearest neighbor site to be vacant so that they can make a thermally activated lattice site hop over those barriers.
Once formed, these defects take a random walk allowing nuclei to move to neighboring lattice sites producing long, correlated cascades. This bulk diffusion process is strongly temperature dependent, with the defect formation rate and defect migration rate following an Eyring law. 

In neutron star crusts, where the thermal energy is typically much lower than the typical Coulomb energy, bulk diffusion within the crystalline domains quenches. However, the structure of GBs allow for increased mobility due to the local disorder and lower coordination of nuclei. These effects lower the activation barriers for hopping, enabling much faster diffusion along GBs than in the crystalline interior. Thus, in cold and highly coupled Yukawa systems relevant to neutron stars, we expect GB diffusion to become the dominant transport mechanism, especially over astrophysically relevant timescales. 

The enhancement of diffusion at GBs has been studied extensively for a long time in the material science literature. In the Fisher model, for example, the GB is treated as a thin film of high diffusivity between two bulk crystals, with diffusion obeying Fick's second law of diffusion in 2D in the interface \cite{fisher1951calculation,Fick}. In this work, we apply this model to GBs in astrophysical Yukawa crystals that are expected in neutron star crusts and white dwarf cores \cite{mestel1952theory,vanHorn1968,Tremblay2019,Saumon2022CurrentChallenges,Blouin2024fundamentals,Ruderman1968,Chamel2008Crusts,caplan2017colloquium}. Following from the Fisher model, one could argue that diffusion in GBs in Yukawa crystals could be modeled as a thin film one unit cell thick where the diffusion coefficient of the nuclei is that of a supercooled liquid. At $\Gamma=350$, about half the melting temperature, the dimensionless diffusion coefficient $D^* = D / \omega_p a_i^2$ of a supercooled liquid predicted by Caplan et al. \cite{CaplanBauerFreeman} is of order $D^*  = 10^{-4}$, which is accessible on MD timescales. In contrast, the fit for the bulk solid obtained in CY24 predicts a diffusion coefficient closer to $D^*  = 10^{-17}$. It is therefore prohibitive to simulate long enough to resolve even one diffusive hop in bulk solids, but if GB diffusion coefficients are comparable to supercooled liquids they should be resolvable with MD, and it is in fact convenient to do so because one should expect bulk diffusion to be completely frozen out.

There are many reasons to think neutron star crusts are polycrystals \cite{horowitz2009breaking,Kobyakov2014,Kobyakov2015,Caplan2018Poly,Caplan2020structure,Baiko2024}. Grains, and by extension GBs, are required essentially by definition if the crystal is not a monocrystal and a monocrystal of stellar scales strains credulity.  On stellar scales, the density varies with depth meaning the interparticle spacing varies. This will create stresses which can only be relieved by adjusting the orientation of the lattice, guaranteeing that some form of stacking fault or dislocation must be encountered when descending through the star. A GB is equivalent to a wall of dislocations. Furthermore, compositional domains due to chemical separation occurring in crystallizing mixtures are also likely in accreted neutron star crusts and in white dwarf cores \cite{HorowitzBerryBrown2007,Medin2011CompDrivenCon,Mckinven2016,Caplan2018Poly,Baiko2024}. Capture layers in neutron star crusts likewise have sharp gradients in nuclear composition, and thus internuclear spacing \cite{Chamel2008Crusts}.
Most importantly, neutron star crusts are subject to a variety of magnetic and elastic forces as the star spins down and as the magnetic field evolves, with starquakes providing a natural mechanism to break the crust and introduce defects including domains \cite{jones2003nature,Thompson2017,Bransgrove_2020, Bransgrove_2025, QuBransgrove2025}. Thus, even with astronomically long times to anneal, there is reason to think that neutron star crusts regularly experience perturbations to their crystalline structure. 

Molecular dynamics (MD) simulations calculate ion trajectories and, by extension, transport properties from first principles. This is especially valuable because we are interested in rare thermally activated diffusive events and the MD simulations probe the thermodynamic distributions naturally. In past MD simulations of Yukawa crystals, GBs and polycrystals have been observed to form naturally, but until now GB diffusion rates have not been rigorously quantified for astrophysics applications, such as neutron star crusts \cite{Hughto2011Diffusion,Caplan2018Poly,Caplan2020structure}.

We are therefore motivated in this work to quantify GB diffusion coefficients in Yukawa crystals with MD simulations. Our MD formalism is introduced in Sec. \ref{sec:meth}, we report on our simulations and results in Sec. \ref{sec:res}, and conclude in Sec. \ref{sec:sum}. 

\section{Methods}\label{sec:meth}

Our molecular dynamics methods using LAMMPS all follow from CY24, but with notable differences in the crystal configurations to create the grain boundaries \cite{LAMMPS}. 

\subsection{Formalism}\label{sec:form}

We calculate two-body interactions between nuclei using the standard repulsive Coulomb potential with a Yukawa screening,
\begin{equation}
V(r_{ij}) = \frac{e^{2} Z_{i} Z_{j}}{r_{ij}} \exp\left(-\frac{r_{ij}}{\lambda}\right),
\end{equation}
\noindent where $r_{ij}$ is the separation between nuclei with charges $eZ_{i}$ and $eZ_{j}$, and $\lambda$ is the electron screening length taken from the relativistic Thomas–Fermi approximation for degenerate electrons,
\begin{equation}
\lambda=\frac{\pi^{1/2}}{(4\alpha k_F)^{1/2}(k_F^2+m_e^2)^{1/4}} \sim \frac{1}{2k_F}\sqrt{\frac{\pi}{\alpha}},
\label{eq.lambda}
\end{equation}
where \( k_F = (3\pi^2 n_e)^{1/3} \) is the electron Fermi momentum, \( m_e \) is the electron mass, and \( \alpha \) is the fine-structure constant, with $n_e = \langle Z\rangle n_i$ with $n_e$ ($n_i$) the number density of electrons (ions). The relativistic approximation is good to within a few percent starting at densities of about $10^8 \ \rm{g/cm^3}$. This is comparable to the cores of the most massive white dwarfs and the outer crusts of neutron stars where one finds the first electron capture layers.

We define the dimensionless screening parameter
\begin{equation}
\kappa = \frac{a_i}{\lambda}.    
\end{equation}
\noindent with ion Wigner-Seitz radius $a_i = (4\pi n_i/3)^{-1/3}$. One can show from Eq. \ref{eq.lambda} that in the limit of relativistic electrons
${\kappa =  18^{1/3} \alpha^{1/2} \pi^{1/6} \langle Z \rangle^{1/3} }$
or ${\kappa(Z) = 0.185 \langle Z \rangle^{1/3}}$. As a lower limit, $\kappa(6) = 0.33$ for carbon in white dwarf cores. As an upper limit, $\kappa(52)=0.69$, corresponding to tellurium at the endpoint of the rp-process in accreted neutron star crusts and is approximately the maximum $Z$ throughout cold catalyzed crusts \cite{schatz2001end}. Accreted crusts may reach as low as $Z=10$ or $12$ just above pycnonuclear capture layers \cite{HaenselZdunik2003,Chamel2008Crusts}. In this work we will simulate with $\kappa=0.333$ and $0.666$ for simplicity, and for ease of comparison with our earlier work which calculated diffusion coefficients for these exact values for the liquid and solid OCP \cite{CaplanBauerFreeman,Caplan2024}.

In this work we consider a pure composition that would be expected for cold catalyzed neutron star crusts and is a good approximation for densities above neutron drip in accreted crusts. In the one-component plasma (OCP), all ions have identical charge $Z$ and mass $m$. The OCP is parametrized by two dimensionless quantities. The first is dimensionless screening $\kappa$ given above and the second is the coupling parameter

\begin{equation}\label{eq:gamma}
    \Gamma = \frac{e^{2} Z^{2}}{a_i T},
\end{equation}
where $T$ is temperature.
In the unscreened limit ($\kappa = 0$), the OCP crystallizes at a critical coupling $\Gamma_{\rm crit} \approx 175.7$, above (below) which the system is solid (liquid) \cite{baiko2022ab}. 

This phase transition moves to lower temperatures (higher $\Gamma$) in the presence of screening. The melt line in $\Gamma \textendash \kappa$ space has been obtained by an analytic fit to molecular dynamics data,
\begin{equation}\label{eq:GammaM}
\Gamma_{\rm M}(\kappa) \approx \Gamma_{\rm crit}\,\frac{e^{c\kappa}}{1 + c \kappa + \frac{1}{2}(c\kappa)^{2}}, \ c \equiv (4\pi/3)^{1/3}
\end{equation}
\noindent where the factor of $c=1.612$ is needed to convert from the convention where $\kappa = n_i^{1/3}/\lambda$  \cite{vaulina2000scaling,Vaulina2002Universal,Silvestri2019}. This fit for the melt line is accurate to one percent for $\kappa \lesssim 1.0$  \cite{vaulina2000scaling}. The correction factor is small, for $\kappa = 0.666$ ($\kappa=0.333$) it is equal to 1.104 (1.018), typical for neutron stars (white dwarfs).

It is convenient to define a modified coupling parameter such that,
\begin{equation}
    \Gamma^{*} = \frac{\Gamma}{\Gamma_{\rm M}(\kappa)}
\end{equation}
so that solid–liquid coexistence universally occurs near $\Gamma^{*}\approx 1$ across all $\kappa$ \cite{Vaulina2002Universal,Caplan2024}. The natural timescale in these systems is the inverse ion plasma oscillation frequency, $\omega_p^{-1}$, where 
\begin{equation}\label{eq:wp}
    \omega_p = \sqrt{\frac{4\pi e^{2} Z^{2} n_i}{m}}.
\end{equation}
\noindent Screening softens the potential and lengthens this oscillation timescale by a factor of $(1+c\kappa +  c^2 \kappa^2 /2)^{1/2}\exp[-ck/2]$ in the dust lattice wave approximation \cite{vaulina2000scaling}. This correction is small for neutron star crust conditions so we take our molecular dynamics timestep to be $dt = \omega_p^{-1}/17$ without adjustment for the screening correction.

\subsection{Initial Configurations}\label{sec:config}

Initial conditions are significantly more detailed than those in CY24 and are generated using the Atomsk code \cite{Caplan2024,hirel2015atomsk}.

To create GBs, we first create several crystal orientations that tile a simulation box and then join them on a side.  There are effectively an infinite number of GBs possible, but number theory imposes helpful constraints on the useful GBs we can fit in a finite volume. In order for the simulation to maintain strict periodicity, the crystallites on either side of the GB must tile the plane without introducing additional artificial GBs due to misalignment across periodic boundaries. This requirement imposes a discrete set of allowed misorientations; only those for which the rotated lattices are commensurate with the simulation box of the chosen size are permitted. As a result, not all GB orientations fit in a periodic box of finite size. This constraint serves a practical purpose: it discretizes the otherwise infinite continuum of possible GB orientations, reducing the problem to a finite set of tractable cases that we take as representative.

\begin{figure}[t]
    \centering
    \includegraphics[width=0.45\textwidth]{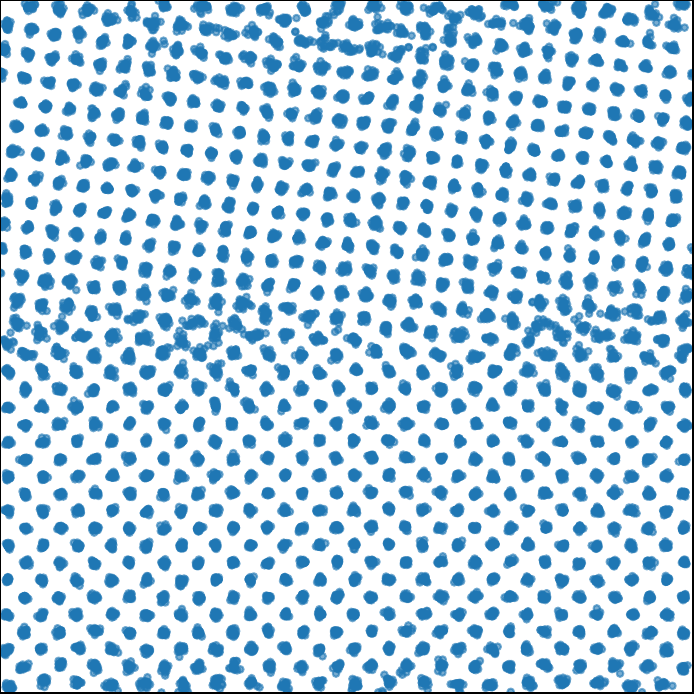}
    \caption{ \textbf{Grain Boundary:} Initial MD configuration for a pure tilt GB viewed along the $xz$-plane, constructed from crystal A (bottom) and crystal B (top). There are two GBs, one in the middle and another at the top due to the periodic boundary.}
    \label{fig:xz_projection}
\end{figure}

To explore the effect of GB orientation on diffusion, three distinct crystal orientations were constructed, which we denote as Crystals A, B, and C in Tab. \ref{tab:crystal_faces}, and we show an example of one GB in Fig. \ref{fig:xz_projection}. Crystal A is the `perfect' body-centered cubic crystal (bcc) that is aligned with the box from CY24, described above, and should be considered our fiducial `reference' configuration. For completeness, it is constructed by tiling the conventional bcc unit cell such that the lattice vectors are aligned with the Cartesian coordinate axes. In this naive setup, the cube-shaped simulation box is periodic in all directions, and the equivalent $(100)$, $(010)$, and $(001)$ planes are aligned with the $x$-, $y$-, and $z$-faces of the box, respectively.

\begin{table}[h]
\centering
\caption{Crystal face summary, with rotation angle about the $z$ axis, and Miller indices sorted by normal to $x$, $y$, and $z$ axes. }
\begin{tabular}{ccc}
\hline
Crystal & $xy$-Rotation & Miller Indices \\
\hline
A & $0^\circ$ & $(100)$, $(010)$, $(001)$ \\
B\textsuperscript{\dag} & $36.87^\circ$ & $(430)$, $(\bar{3}40)$, $(001)$ \\
C\textsuperscript{\ddag} & $16.62^\circ$ & $(24 \ 7 \ 0)$, $(\bar{7} \ 24 \ 0)$, $(001)$  \\
\hline
\end{tabular}
\begin{flushleft}
\textsuperscript{\dag} $\tan^{-1}(3/4)$ for $\Sigma5$ CSL \\
\textsuperscript{\ddag} $\tan^{-1}(7/24)$ for  $\Sigma25$ CSL
\end{flushleft}

\label{tab:crystal_faces}
\end{table}

Crystal B is generated by applying a rigid-body rotation of $\tan^{-1}(3/4) \approx 36.87^\circ$ about the $z$-axis to the reference lattice. This rotation maps the $x$- and $y$-normals of the reference lattice to directions normal to the $(430)$ and $(\bar{3}40)$ planes, respectively, while the $xy$-plane presents a rotated $(001)$ face. The specific angle is selected to ensure exact tiling of the periodic box, exploiting the Pythagorean triple $(3,4,5)$ which guarantees that the rotated lattice vectors still match the periodicity of a box face. Note that this is the rotation for the common $\Sigma5$ GB, specifically, it corresponds to a misorientation of approximately $36.87^\circ$ about the $z$-axis, which produces a coincidence site lattice (CSL) where 1 in every 5 lattice sites overlap. 

Crystal C is constructed similarly, with a rotation to the reference lattice of $\tan^{-1}\left(7/24 \right) \approx 16.26^\circ$ about the $z$-axis. This transformation leaves the $(001)$ plane unchanged except for its in-plane rotation, but it transforms the side faces into  $(24 \ 7 \ 0)$ and $(\bar{7} \ 24 \ 0)$. This orientation corresponds to a $\Sigma25$ coincidence site lattice obtained from the $(7,24,25)$ Pythagorean triple. To maintain strict periodicity in this case, the simulation box must contain 25 unit cells along at least one direction (e.g., the $x$-axis) to ensure that both the rotated and unrotated lattices fit precisely within the box. 

We prepare 11 unique GBs from these crystals. To construct them, the three crystals are split midway along an axis and the two halves joined create a misorientation between the grains. For consistency, the resulting bicrystals are rotated so that GBs are orthogonal to the $z$-axis. 
This procedure introduces two GBs: one located at the center of the simulation box ($z = L_z/2$), and one due to periodic boundary conditions at the top and bottom ($z = 0$ and $z = L_z$). After initial construction, the systems were energy-minimized and equilibrated at finite temperature using the NVT ensemble to relieve residual stresses. 
This step also allows the GB to minimize energy without finding the $\gamma-$surface, as the periodicity in $x$ and $y$ allow the grains to displace by a fraction of a lattice spacing to find a minimum energy alignment. 
In all cases, the simulation box contains between $20 \times 20 \times 20$ unit cells and $25 \times 25 \times 20$ unit cells, with the larger boxes required for Crystal C.

%\begin{table}[t]
%\centering
%\caption{Unique grain boundaries considered in this work.}
%\begin{tabular}{ccccc}
%\hline
%\# & 
%\makecell{Crystal \\ Pair} & 
%\makecell{Crystal 1 \\ Face} & 
%\makecell{Crystal 2 \\ Face} & 
%\makecell{Boundary \\ Type} \\
%\hline
%$1$  & A - B & $(100)$ & $(100)$ & Twist \\
%$2$  & A - B & $(100)$ & $(430)$ & Tilt \\
%$3$  & A - C & $(100)$ & $(100)$ & Twist \\
%$4$  & A - C & $(100)$ & $(24\ 7\ 0)$ & Tilt \\
%$5$  & B - B & $(100)$ & $(430)$ & Tilt \\
%$6$  & B - C & $(100)$ & $(100)$ & Twist \\
%$7$  & B - C & $(100)$ & $(24\ 7\ 0)$ & Tilt \\
%$8$  & B - C & $(430)$ & $(100)$ & Tilt \\
%$9$  & B - C & $(430)$ & $(24\ 7\ 0)$ & Tilt \\
%$10$ & C - C & $(100)$ & $(24\ 7\ 0)$ & Tilt \\
%$11$ & B - B & $(430)$ & $(340) $ & TWist \\
%\hline
%\end{tabular}
%\label{tab:uniquegbs}
%\end{table}

We classify the GBs as either tilt, twist, or mixed depending on the crystallographic orientation of the adjoining grains and the boundary plane. In a twist boundary, the two crystals are misoriented by a rotation about an axis that is normal to the boundary plane typically resulting in a symmetric, planar interface with relatively low energy and more stable atomic arrangements at the interface. This symmetry and stability makes these boundaries less prone to migration, as they are unlikely to induce asymmetric stresses in the crystallites that would cause one to be energetically favored over the other, and thus grow by GB migration until the interfaces meet and the box is annealed to just one crystal orientation. 
In contrast, tilt boundaries involve a rotation about an axis parallel to the boundary plane. These interfaces often display more complex structures, including `jagged' dislocation arrays or steps (like in Fig. \ref{fig:xz_projection}), generally making them more susceptible to dynamic migration due to thermal fluctuations or internal stresses. Our preliminary tests find that temperatures below about half of melting suffice to freeze out GB migration, eliminating the need for particle tagging, frozen layers, or any more sophisticated tricks tracking boundaries for isolating GB diffusion coefficients.

\section{Simulations}\label{sec:res}

\subsection{Fiducial Runs}

\begin{figure*}[p]
    \centering
    \includegraphics[width=0.85\textwidth]{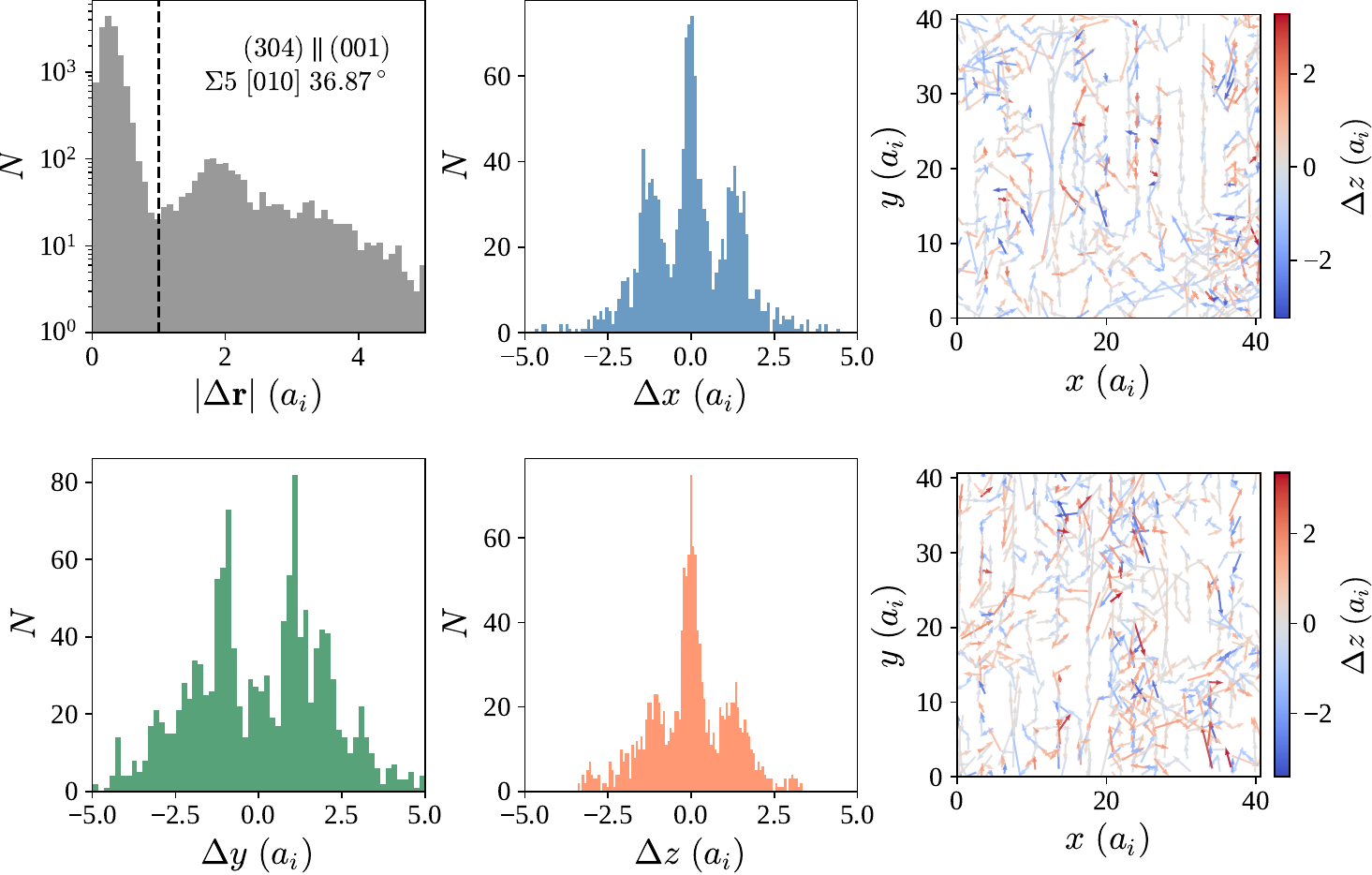}
    \vspace{1mm}
    \hrule
    \vspace{5mm}
    \includegraphics[width=0.85\textwidth]{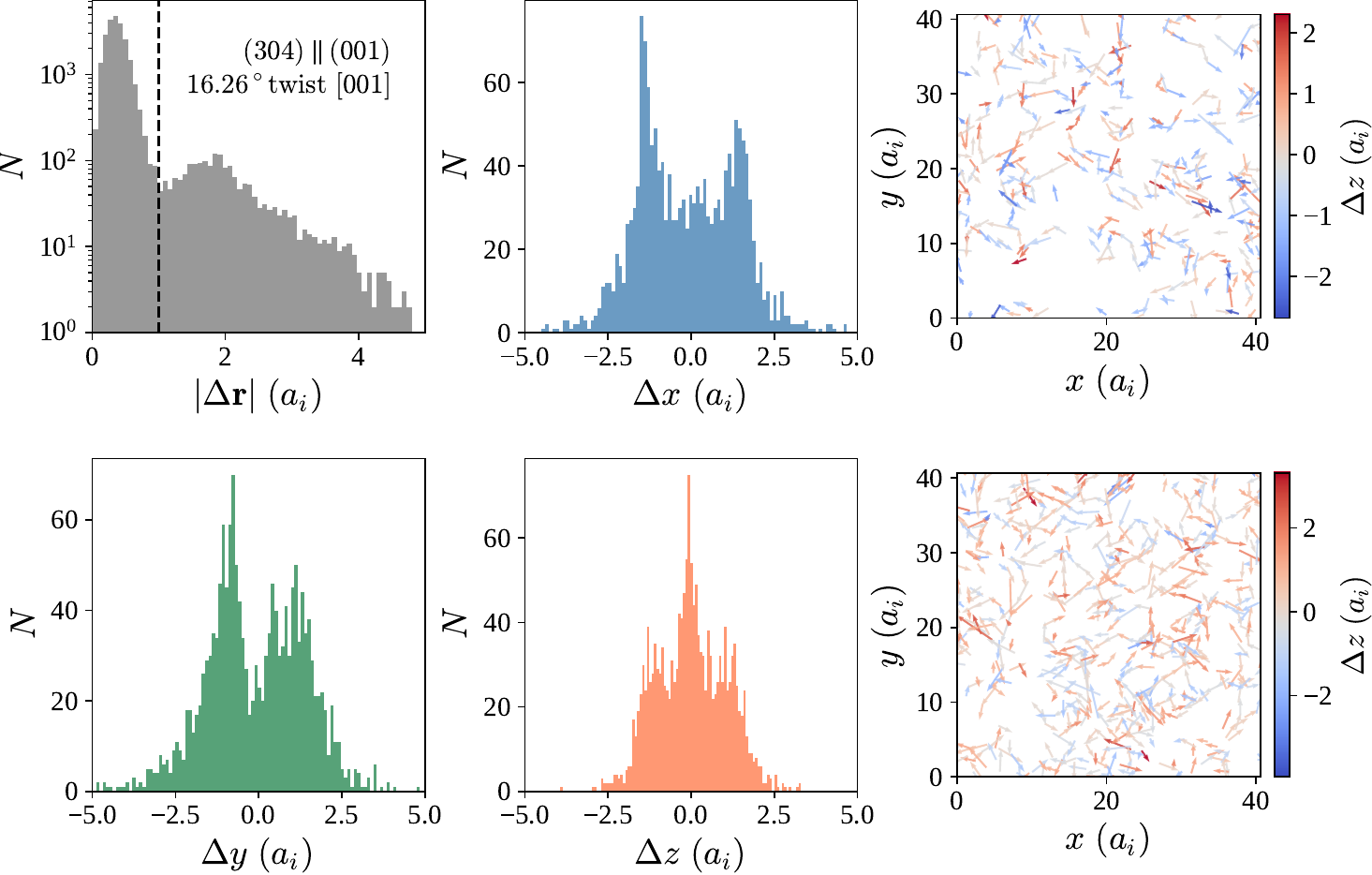}
    \caption{ \textbf{Diffusion analysis of the fiducial runs.} We simulate the pure tilt $\Sigma5$ boundary (top) and a mixed boundary obtained by twisting the (001) crystal (bottom). (Top left) A histogram of displacement magnitudes $|\Delta \mathbf{r}|$ of all particles in the simulation show that about a thousand nuclei diffuse more than one $a_i$, consistent with the expectation that about a tenth of the particles are in a GB, and no diffusive hops are observed outside of a GB. For the thousand particles that diffuse by more than a lattice spacing, we show the displacement in (top center) $\Delta x$, (bottom left) $\Delta y$, and $\Delta z$ (bottom center). It is apparent that there are abundant diffusive steps in the $x$ and $y$ directions, but only incidental diffusion in $z$. To visualize diffusion at the two GBs, we plot displacement vectors for the particles that move (right). In the pure tilt crystal, diffusion in the $y$ direction dominates at both the top GB (top right) and central GB (bottom right) because particles move in the `ridges' of crystal B; this is the into-the-page direction of Fig. \ref{fig:xz_projection}, while the twist destroys the $y$-axis symmetry. Few particles move vertically by more than one unit cell length (vector color). 
    }
    \label{fig:fiducial_sixpanel}
\end{figure*}

We begin by characterizing two related GBs in detail as fiducial examples. 

In Fig. \ref{fig:fiducial_sixpanel} we characterize the $(403) \parallel (001)$ GB diffusion for a pure tilt boundary (top) and mixed tilt-twist boundary (bottom). The pure tilt case, which is the high symmetry $\Sigma5$ boundary constructed from crystals A and B, is shown in Fig. \ref{fig:xz_projection}. The mixed boundary is constructed similarly from crystal B and C, so that in addition to the tilt there is an additional rotation of $\tan^{-1}(7/24)$ about the $[001]$ axis, equivalent to rotating the `bottom' crystal in Fig. \ref{fig:xz_projection} about the vertical axis. Both simulations in Fig. \ref{fig:fiducial_sixpanel} were run at $\Gamma=450$ and $\kappa = 0.666 $. The configuration in Fig. \ref{fig:xz_projection} was obtained by raising the initial crystal to the desired temperature of $\Gamma=450$ by evolving the initial configuration for $3\,000$ timesteps in the NVT ensemble and rescaling the velocities every 10 timesteps to the desired temperature. In both simulations there are two GBs, one in the middle and one at the top and bottom from the periodic boundary. These boundaries drift slightly at early times to minimize energy during this equilibration. They were then evolved for $10^6$ timesteps at constant energy for data collection. 

Consider the top panel in Fig. \ref{fig:fiducial_sixpanel}, with a pure tilt interface. With a box size of $20\times20\times20$ unit cells, the Fisher model would predict that the volume occupied by the two GBs is $2\times20\times20$ unit cells, containing $1\, 600$ particles or a tenth of all nuclei. We do not resort to other techniques, like fixing the particles in the bulk, because at low screenings the Yukawa plasma has long range correlations that are important to the dynamics. In Fig. \ref{fig:fiducial_sixpanel}, the number of particles that diffuse by at least $a_i$ (top left) is a useful proxy for the number of particles that have experienced at least one diffusive hop, and is about $1\,000$, consistent with our argument. At this high temperature there is some small uncertainty due to the thermal noise, but the error is at most a few tens of particles here. At lower temperatures the separation between the first and second peak is very clear. 

We isolate these particles for further analysis, and show histograms of their displacement components in $\Delta x$, $\Delta y$, and $\Delta z$. While $\Delta x$ and $\Delta y$ show large numbers of particles displaced by at least $a_i$ in pairs of peaks, the central peaks differ starkly and show significantly more diffusive hops in $y$ than in $x$. Consistent with our expectation that diffusive hops are isolated to the GB,  $\Delta z$ shows only weak secondary peaks. Furthermore, the spread in $\Delta z$ does not extend beyond about $2.5 a_i$, lacking the extended tails seen for $\Delta x$ and $\Delta y$, so the displacements orthogonal to the GBs are comparable to the expected GB thickness. 

No diffusive hops ($|\Delta \mathbf{r}| > a_i$) are observed in the bulk, making it trivially easy to study the two GBs separately. We visualize the diffusive activity in the GB at the top of the box (top right) and the center of the box (bottom right) and find that the majority of the diffusion is in the $y$ direction in long diffusive cascades, likely driven by vacancies or interstitials migrating along the `groves' or `ridges' of the uneven interface of the tilted crystal. 
Note that this sort of behavior is known for terrestrial materials in the literature, and this mechanism goes by many other names such as point-defect avalanches \cite[\textit{e.g.} ][]{chesser2022point}. It is possible that these cascades are crowdions, quasi-particles where $N$ particles occupy $N\pm1$ co-linear lattice sites. Such crowdions would likely have lower activation energies for individual hops, enabling the rapid diffusion of many particles. 

We now consider mixed twist-tilt interface in the bottom panel of Fig. \ref{fig:fiducial_sixpanel}. The differences with the pure tilt boundary are immediately apparent. With the loss of the $y-$axis symmetry, there is little difference in the distribution of $\Delta x$ and $\Delta y$. The spread in $\Delta z$ is slightly broader though still symmetric, suggesting that the GB is not migrating. When comparing the central interface and the top interfaces, the central interface appears slightly more active during our simulation. This simulation requires a larger number of particles, and is run in a box of $25\times25\times20$ unit cells, so the interfaces contain $2\,500$ particles. The temperature only has fluctuations of order $10^{-2}$ during the fiducial runs so we expect the uncertainty in the diffusion coefficients we calculate to be dominated by stochasticity in the small finite number of lattice site hops. 

As expected, diffusion in GBs in Yukawa crystals is highly sensitive to the exact crystal orientation at GBs, similar to what is known for grains in terrestrial materials. 

\subsection{Grain Boundary Diffusion Coefficients}\label{sec:GBDCs}

For 11 unique GBs we run simulations over a range of temperatures spanning $450 \leq \Gamma \leq 1100$.
For every boundary and $\Gamma$ simulated we run sufficiently long so that at least 100 particles that have displaced by $|\Delta\mathbf{r}|>a_i$, typically $10^6$ timesteps is sufficient for $\Gamma \gtrsim 800$. Our longest simulations are for the least mobile GBs at highest $\Gamma$ and are no more than $10^7$ timesteps. 
At high temperatures, we find that the number of particles moving at least $a_i$ reaches a maximum approximately equal to the number of particles in a cross sectional slice of the box two unit cells thick, consistent with our expectation that diffusion is confined to our two interfaces that are each one unit cell thick.

\begin{figure*}[t]
    \centering
    \includegraphics[width=0.8\linewidth]{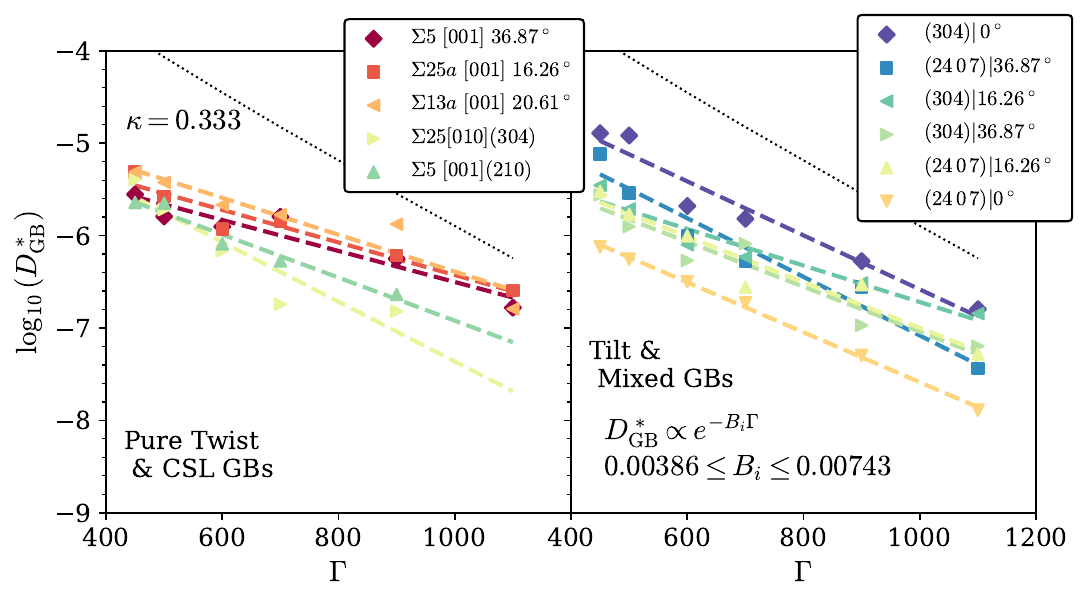}
    \includegraphics[width=0.8\linewidth]{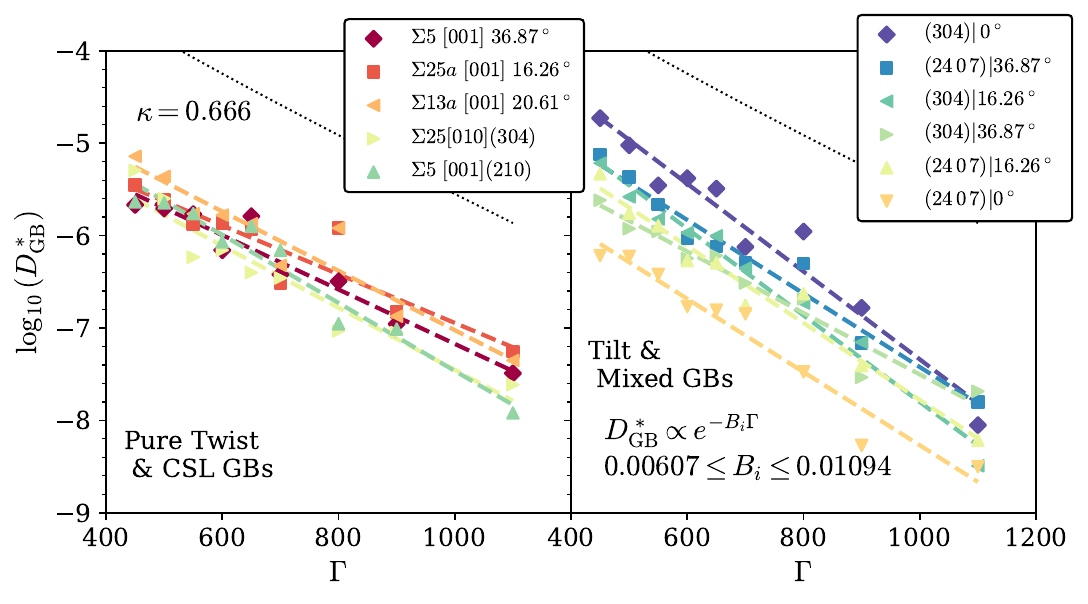}
    \caption{ \textbf{Grain Boundary Diffusion Coefficients:} Diffusion coefficients for $\kappa=0.333$ (top) and $\kappa=0.666$ (bottom) for pure twist and CSL GBs (left) and tilt and mixed GBs (right). GB diffusion coefficients consistently show only weak sensitivity to the specific crystal orientation. The Eyring model slope is generally steeper for tilt and mixed GBs than pure twist and high symmetry GBs. The fit for a liquid (dotted) from Caplan et al. \cite{CaplanBauerFreeman} has been extrapolated to supercooled temperatures, and shows that GB diffusion follows approximately the same Eyring model slope but rescaled down by one to two orders of magnitude.}
    \label{fig:DGBk0.666}
\end{figure*}

\begin{table*}[]
\setlength{\tabcolsep}{10pt} 
\renewcommand{\arraystretch}{1.2} 
\centering
\caption{Eyring parameters $\alpha$ and $\beta$ for boundary types at $\kappa_1 = 0.333$ and $\kappa_2 = 0.666$. Best fit $A(\kappa) \equiv \alpha \Gamma_\mathrm{M}(\kappa)$ and $B(\kappa) \equiv \beta / \Gamma_\mathrm{M}(\kappa)$ in Fig. \ref{fig:DGBk0.666} can be recovered using $\Gamma_\mathrm{M}(0.333)=178.8$ and  $\Gamma_\mathrm{M}(0.666)=194$. }
\begin{tabular}{lcc|cc|cc}
\toprule
\textbf{} 
& \multicolumn{2}{c|}{\(\kappa_1 = 0.333\)} 
& \multicolumn{2}{c|}{\(\kappa_2 = 0.666\)} 
& \multicolumn{2}{c}{Ratios} \\
& \(\alpha_1\) & \(\beta_1\)
& \(\alpha_2\) & \(\beta_2\)
& \(\alpha_1 / \alpha_2\) & \(\beta_1 / \beta_2\) \\
\midrule
\multicolumn{7}{l}{\textbf{Twist \& CSL Grain Boundaries}} \\
$\Sigma5\ [001]\ 36.87^\circ$          & \(8.39 \times 10^{-8}\)  & 0.6902 & \(3.04 \times 10^{-7}\)  & 1.3153 & 0.28 & 0.5247 \\
$\Sigma25a\ [001]\ 16.26^\circ$          & \(1.21 \times 10^{-7}\)  & 0.7259 & \(2.52 \times 10^{-7}\)  & 1.1776 & 0.48 & 0.6165 \\
$\Sigma13a\ [001]\ 20.61^\circ$          & \(2.26 \times 10^{-7}\)  & 0.8207 & \(7.96 \times 10^{-7}\)  & 1.4356 & 0.28 & 0.5717 \\
$\Sigma25[010](304)$            & \(4.11 \times 10^{-7}\)  & 1.3285 & \(4.14 \times 10^{-7}\)  & 1.4996 & 0.99 & 0.8859 \\
$\Sigma5\ [001](210)$         & \(1.44 \times 10^{-8}\)  & 0.9602 & \(8.48 \times 10^{-7}\)  & 1.6432 & 0.17 & 0.5843 \\
\midrule
\multicolumn{7}{l}{\textbf{Tilt \& Mixed Grain Boundaries}} \\
$(304) | \, 0^\circ$       & \(1.23 \times 10^{-6}\)  & 1.2033 & \(1.33 \times 10^{-5}\)  & 2.1223 & 0.09 & 0.5670 \\
$(24 \, 0 \, 7) | 36.87^\circ$      & \(6.90 \times 10^{-7}\)  & 1.3052 & \(1.85 \times 10^{-6}\)  & 1.7770 & 0.37 & 0.7345 \\
$(304) |  16.26^\circ$       & \(1.01 \times 10^{-7}\)  & 0.8135 & \(3.99 \times 10^{-6}\)  & 2.0952 & 0.03 & 0.3883 \\
$(304) | 36.87^\circ$       & \(1.38 \times 10^{-7}\)  & 1.0013 & \(3.44 \times 10^{-7}\)  & 1.4860 & 0.40 & 0.6738 \\
$(24 \, 0\, 7)  | 16.26^\circ$      & \(1.66 \times 10^{-7}\)  & 1.0174 & \(1.25 \times 10^{-6}\)  & 1.8585 & 0.13 & 0.5474 \\
$(24 \,0 \,7) |  0^\circ$      & \(7.31 \times 10^{-8}\)  & 1.1121 & \(2.54 \times 10^{-7}\)  & 1.7693 & 0.29 & 0.6286 \\
\bottomrule
\end{tabular}
\label{tab:alpha_beta}
\end{table*}

Grain boundary diffusion coefficients were computed from the mean squared displacement (MSD) of atoms in the identified GB regions. Our methods follow from our past work \cite{Hughto2011Diffusion,Caplan2024,CaplanYaacoub2025}. For each atom $i$, the displacement relative to its initial position was computed and the MSD was averaged over the number of nuclei in the grain boundary $N_{\text{GB}}$, taken to be one tenth of the particles in the box. This is because our boxes are all 20 unit cells thick in $z$, so that the two interfaces together constitute a total interface thickness 2 unit cells thick. In general, 

\begin{equation}
\mathrm{MSD}(t) = \frac{1}{N_{\text{GB}}} \sum_{i=1}^{N_{\text{GB}}}  \left\lvert \vec r_i(t) - \vec r_i(0) \right\rvert^2 .
\end{equation}

\noindent Because GB diffusion is primarily confined to the in-plane directions ($x$, $y$), the MSD was projected accordingly:

\begin{equation}
\mathrm{MSD}_{\parallel}(t) = \frac{1}{N_{\text{GB}}} \sum_{i=1}^{N_{\text{GB}}}  (x_i(t)-x_i(0))^2 + (y_i(t)-y_i(0))^2 
\end{equation}

\noindent where the parallel `{$\parallel$}' subscript is used to denote in-plane behavior and helps to differentiate any quantities here from those calculated for bulk diffusion coefficients reported in CY24.

To eliminate noise from particles undergoing only small thermal vibrations near their lattice sites, we use the Heaviside step function \(\Theta(x)\) to filter out particles whose displacement remains within a cutoff radius $R_c = a_i$. We further define the in-plane MSD as:

\begin{equation}
\mathrm{MSD}_{\parallel}(t) = \left\langle \Theta\left( \left| \mathbf{r}_j(t) - \mathbf{r}_j(0) \right| - R_c \right) \left| \mathbf{r}_j(t) - \mathbf{r}_j(0) \right|^2 \right\rangle .
\end{equation}

\noindent  Assuming Fickian behavior, the in-plane diffusion coefficient was obtained from the linear region of this MSD:

\begin{equation}
D_{\mathrm{GB}} = \frac{1}{4}\,\frac{d}{dt} \mathrm{MSD}_{\parallel}(t).
\end{equation}

%\begin{equation}
%D_{\parallel}(T) = D_0 \exp\left(-\frac{Q}{k_B T}\right),
%\end{equation}

%where $D_0$ is the pre-exponential factor, $Q$ is the activation energy for diffusion, $k_B$ is Boltzmann’s constant, and $T$ is the temperature in Kelvin. Recognizing that $Q$ is set by the Coulomb energy $e^2 Z^2/a_i$ times some prefactor, this can all be rewritten in a suitable dimensionless form following from CY24, 

\noindent This is made dimensionless by $D^*_{\mathrm{GB}} \equiv D_{\mathrm{GB}} / a_i^2 \omega_p$. We take our Eyring model for diffusion to be that of CY24, 

\begin{equation}\label{eq:Dstar}
    D^*_{\mathrm{GB}} = \frac{A_{}(\kappa)}{\Gamma} e^{-B_{}(\kappa) \Gamma} = \frac{\alpha}{\Gamma^{*}} e^{-\beta \Gamma^{*} }
\end{equation}

\noindent with $A_{} \equiv \alpha_{} \Gamma_{\rm M}(\kappa)$ and $B_{} \equiv \beta_{} / \Gamma_{\rm M}(\kappa)$ and $\Gamma_{\rm M}$ from Eq. \ref{eq:GammaM}. At our highest temperatures $\mathrm{MSD}_{\parallel}(t)$ is nearly linear, as for a liquid, though at our lowest temperatures the stochastic nature of the hops becomes apparent and $\mathrm{MSD}_{\parallel}(t)$ resembles the `stairstep' in Fig 1 of CY24. 

In Fig. \ref{fig:DGBk0.666} we present $D^*_\mathrm{GB}$ calculated from our MD simulations and their best fit lines for $\kappa=0.333$ and $\kappa=0.666$. 
Trends with GB orientation are generally consistent between each $\kappa$, for example $D^*_\mathrm{GB}$ is highest for the $(304)|0^{\circ}$ tilt interface and slowest in the $(24 \ 0 \ 7)|0^{\circ}$ tilt interface. The only notable difference is observed for the pure twist and CSL GBs. While the spread is quite small among GBs studied at $\kappa =0.666$, it is somewhat larger at $\kappa=0.333$. We suspect this is more likely than not stochastic. We simulate at higher resolution in $\Gamma$ for $\kappa=0.666$, and so the fits to the $\kappa=0.333$ simulations can be more heavily biased by a few points. The presence (or absence) of one or two especially prolific avalanches can skew our calculations of $D^*_\mathrm{GB}$ at any given $\Gamma$,  especially at high $\Gamma$ when diffusion is slow. 
In all cases, we find that diffusion in GBs is slower than in an equivalent supercooled liquid by one to two orders of magnitude (dotted line in Fig. \ref{fig:DGBk0.666}), but the Eyring model slope $B_i$ is generally consistent with the best fits for a liquid from \cite{CaplanBauerFreeman}. This is strongly supportive of interpretations of GBs in Yukawa crystals as thin films of liquid. 

We now consider if GB diffusion exhibits the same kind of universal scaling observed for the bulk solid in CY24. In that work, we found that the Eyring model from Eq. \ref{eq:Dstar} holds near melting for all screening lengths using $\alpha = 787$ and $\beta = 20$. In Tab. \ref{tab:alpha_beta} we compute $\alpha$ and $\beta$ from the best fit $A(\kappa)$ and $B(\kappa)$ in Fig. \ref{fig:DGBk0.666}, using ${\Gamma_\mathrm{M}(\kappa=0.333) = 178.8}$ and ${\Gamma_\mathrm{M}(\kappa=0.666) = 194}$ to
normalize by the melting condition for a given $\kappa$. If the universality found in CY24 holds for GBs, we would expect $\alpha(0.333)/\alpha(0.666) \approx1$ and likewise for $\beta$, but this is not what is observed. The best fit $\alpha$ and $\beta$ are noticeably smaller at lower $\kappa$ and these differences are large enough to merit comment. For a nucleus diffusing in a GB the nearest neighbors are the amorphous thin film while the next nearest neighbors are a regular lattice, so the relative importance of the lattice to the barrier landscape of the boundary will depend on the screening. Qualitatively, at higher $\kappa$ the GB will be more strongly dominated by the nearest neighbors and so the interface becomes increasingly liquid-like at higher $\kappa$.

\section{Summary}\label{sec:sum}

From our molecular dynamics simulations of grain boundary diffusion in Yukawa crystals we find that the diffusion coefficients are well fit by a simple Eyring model, as is known for the solid and liquid. Exact values for the diffusion coefficients in GBs are fairly insensitive to the exact orientation of crystallites, allowing future authors to have confidence that our simple fits can reliably describe a wide variety of possible systems.

As a practical matter, we can make a general recommendation that future authors interested in GB diffusion coefficients simply take $D^*$ for a supercooled liquid and reduce it by one to two orders of magnitude. 
This is because the parameters of the best fit Eyring models for $D^*_\mathrm{GB}$ are approximately that for a supercooled liquid. This is well understood with the qualitative physics of the Fisher model for GBs, where GBs are a locally amorphous thin film \cite{fisher1951calculation}. With the understanding that $D^*$ is governed by activation physics that depend on the local landscape of the classical potential surrounding a particle, it is clear that $D^*_\mathrm{GB}$ should be liquid-like and enormously enhanced over that of a bulk solid. Indeed, we conclude that at temperatures relevant for white dwarf cores and neutron star crusts, $D^*_\mathrm{GB}$ should be about 20 orders of magnitude greater than $D^*$ for a bulk solid. This strongly suggests that under realistic conditions in neutron star crusts, GBs may be the dominant source of viscous dissipation and creep, depending on the exact grain size which remains uncertain. The plastic viscosity of solid neutron star matter is an essential input for models of the crust under extreme stress \citep[\textit{e.g.}][]{Beloborodov_2014, Li_2015,Li_2016,Thompson2017,Lander_2019,Gourgouliatos_2021}. However, there are currently large uncertainties in these models because the literature lacks a microscopic theory of viscous dissipation and plastic flow in the crust. 
The present work studied diffusion in an isolated GB, which is clearly a key ingredient for such a theory. Future work
should investigate the formation of complex networks of grain boundaries and the resulting dissipation in crystals under applied strains.

The non-universality is interesting but also not so important for astrophysics. It is reasonable that diffusion in GBs demonstrates some dependence on the exact screening length, but the strength of this effect was not so large to affect our recommendation above.

While there is some spread in our MD data due to the stochastic nature of this problem and the challenge of annealing the initial configurations, this is not a serious concern for astrophysics. The spread in $D^*_\mathrm{GB}$ has been shown to be fairly tight for the region of $\Gamma \textendash \kappa$ space we surveyed that is most relevant for neutron star crusts. Likewise, there is some uncertainty due to other choices in our formalism, for example assuming the GB interfaces are one unit cell thick and neglecting any $z$-displacement due to particles joining the interface from the bulk, but these uncertainties will be smaller than the MD stochasticity that clearly dominates.

% The \nocite command causes all entries in a bibliography to be printed out
% whether or not they are actually referenced in the text. This is appropriate
% for the sample file to show the different styles of references, but authors
% most likely will not want to use it.
%\nocite{*}

\begin{acknowledgments}
The authors thank Jeong Seop Yoon for discussion. 
This work was supported by a grant from the Simons Foundation (MP-SCMPS-00001470) to MC. This research was supported in part by the National Science Foundation under Grant No. NSF PHY-1748958. Financial support for this publication comes from Cottrell Scholar Award \#CS-CSA-2023-139 sponsored by Research Corporation for Science Advancement. %KITP  
MC and AB thank the KITP for hospitality and MC acknowledges support as a KITP Scholar. AB is supported by a PCTS fellowship and a Lyman Spitzer Jr. fellowship. This research was supported in part by Lilly Endowment, Inc., through its support for the Indiana University Pervasive Technology Institute.

\end{acknowledgments}

%\bibliography{bibliography}% Produces the bibliography via BibTeX.

%apsrev4-2.bst 2019-01-14 (MD) hand-edited version of apsrev4-1.bst
%Control: key (0)
%Control: author (8) initials jnrlst
%Control: editor formatted (1) identically to author
%Control: production of article title (0) allowed
%Control: page (0) single
%Control: year (1) truncated
%Control: production of eprint (0) enabled
%

\end{document}